\documentclass[twocolumn,showpacs,superscriptaddress,pra,appendix]{revtex4}

\usepackage{mathrsfs}
\usepackage{amsmath}
\usepackage{amssymb}
\usepackage{graphicx}
\usepackage{dcolumn}
\usepackage{bm,braket}
\usepackage{subfigure}
\usepackage{color}
\usepackage{ifpdf}
\usepackage{epstopdf}
\usepackage{CJK}
\usepackage{float}

\begin{document}

\title{Observation of Quantum Equilibration in Dilute Bose Gases}
\begin{CJK*}{UTF8}{gbsn}

\author{Linxiao Niu(牛临潇)}
      \affiliation{School of Electronics Engineering and Computer Science, Peking University, Beijing 100871, China}
\author{Pengju Tang(唐鹏举)}
      \affiliation{School of Electronics Engineering and Computer Science, Peking University, Beijing 100871, China}
\author{Baoguo Yang(杨保国)}
      \affiliation{School of Electronics Engineering and Computer Science, Peking University, Beijing 100871, China}
\author{Xuzong Chen(陈徐宗)}
      \affiliation{School of Electronics Engineering and Computer Science, Peking University, Beijing 100871, China}
\author{Biao Wu(吴飙)}\email{wubiao@pku.edu.cn}
      \affiliation{International Center for Quantum Materials, School of Physics, Peking University, Beijing 100871, China}
      \affiliation{Collaborative Innovation Center of Quantum Matter, Beijing 100871, China}
      \affiliation{Wilczek Quantum Center, College of Science, Zhejiang University of Technology,  Hangzhou 310014, China}
       \affiliation{Synergetic Innovation Center for Quantum Effects and Applications, Hunan Normal University, Changsha 410081, China}
\author{Xiaoji Zhou(周小计)}\email{xjzhou@pku.edu.cn}
      \affiliation{School of Electronics Engineering and Computer Science, Peking University, Beijing 100871, China}

\begin{abstract}
We investigate experimentally the dynamical relaxation of a
non-integrable quantum many-body system to its equilibrium state. A
Bose-Einstein condensate is loaded into the first excited band of an
optical lattice and  let to evolve up to a few hundreds of milliseconds.
Signs of quantum equilibration are observed.
There is a period of time, roughly 40 ms long, during which both the aspect
ratio of the cloud and its momentum distribution remain constant. In particular,
the momentum distribution has a flat top and is not a Gaussian thermal distribution.
After this period, the cloud becomes classical as its momentum
distribution becomes Gaussian.
\end{abstract}


\maketitle
\end{CJK*}

\section{Introduction}
The second law of thermodynamics states that the entropy of an
isolated system never decreases~\cite{Huang}. When it is applied to
quantum systems,  the second law implies that an isolated quantum
system will dynamically relax to an equilibrium state that has a
maximized entropy. Many physicists including Pauli and Schr\"odinger
had attempted to understand this law quantum
mechanically~\cite{goldstein}. Von Neumann was clearly the most
successful as he proved both quantum ergodic theorem and quantum
H-theorem~\cite{neumann1929,neumann2010}. According to these two
theorems show that most of the non-integrable quantum systems, which
are ubiquitous in nature~\cite{GutzwillerBook}, will indeed relax
dynamically to an equilibrium state, where the macroscopic
observables fluctuate only slightly and the entropy is maximized
with small fluctuations. These two theorems have now been improved
and put in a more transparent framework and on a firmer
footing~\cite{reimann2008,han2014}.

Experimental observation of the dynamical relaxation of an isolated
quantum system had been almost impossible since isolated quantum
systems are very hard to prepare in experiments.  This situation was
changed with the realization of Bose-Einstein condensation (BEC) in
dilute atomic gases~\cite{Anderson}. A BEC in such an experiment has
no physical contact with a heat bath as it is hold either in a
magnetic or an optical trap. As demonstrated in interference and
vortex experiments~\cite{Andrews,Madison} and also in our recent
experiment~\cite{ZhongKai}, a BEC can stay in a pure quantum state
or  the BEC can be regarded as an isolated quantum system up to a
few hundred milliseconds. This shows that it is now experimentally
possible to study the dynamical relaxation of an isolated quantum
system. Such a possibility has generated a great deal of theoretical
interests. Along with many theoretical
works~\cite{Tumulka,Winter2006,linden2009,short2011,Yukalov2011,zhuang2013,zhuang2014},
there have already been experimental studies on this issue. However,
almost all of the experiments are focused on one dimensional
integrable quantum
systems~\cite{Weiss2006,Gring,Smith,Schreiber2015,Bordia2016}. To
the best of our knowledge, the dynamical relaxation of a
non-integrable quantum system was only studied in
Ref.~\cite{Henn2009,Caracanhas2013Self,Seman2010Route} with the
focus on quantum turbulence. As integrable systems are rare and
almost all interacting systems in nature are
non-integrable~\cite{GutzwillerBook}, it appears more important to
study the dynamical relaxation of non-integrable quantum systems.
 \begin{figure}
\begin{center}
 \includegraphics[width=0.4\textwidth]{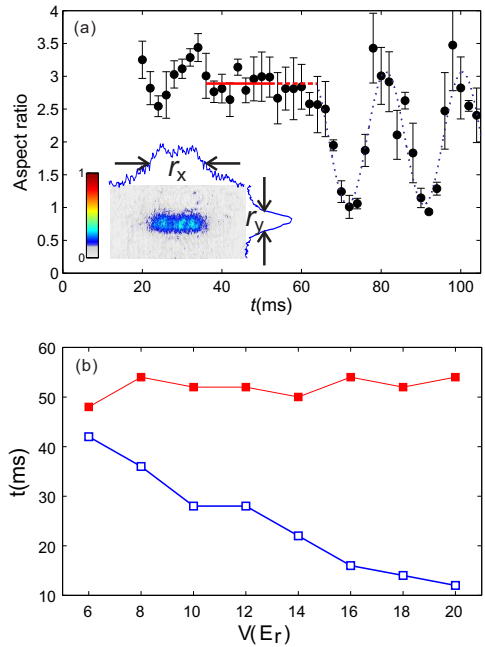}
\end{center}
\caption{(color online) (a) Aspect ratios of the atomic cloud at
different holding times with the lattice strength $V=8E_\mathrm{r}$.
The TOF image at 50 ms is shown in the inset to show how
the aspect ratio is extracted. Each point is the average over five experiments and the
error bar is the standard error. The red solid line indicates the
plateau during which the aspect ratio remains constant while the red dashed line
shows the transition period between two stages. (b) The starting
(blue) and ending times (red) of the plateau  for different lattice
strengths. } \label{figure1}
\end{figure}

In this work we study experimentally the dynamical relaxation of an
isolated non-integrable quantum system. This is achieved by loading  a BEC
into the first excited band of an optical lattice. The BEC is then let
to evolve in time up to 400 ms. The aspect ratio of the BEC cloud after free
expansion is found to oscillate initially and then becomes constant
during a time window between roughly 35 ms and 50 ms (see
Fig.\ref{figure1}), during which a plateau is formed. The length of such a
plateau increases with the strength of optical lattice and is around
40 ms long for an optical lattice of $20E_\mathrm{r}$. During this
time plateau, the overall feature of the BEC cloud remains largely
unchanged. Specifically, the momentum distribution of the cloud
 does not change, and has a flat top such that it can not be fitted with any known
thermal distribution.  Such a plateau strongly indicates that a
quantum equilibrium is reached. After the plateau, the oscillations
in the aspect ratio are resumed with a frequency that is twice of
the trapping frequency. The system eventually reaches the classical
thermal equilibrium, where the momentum distribution of the BEC is
Gaussian. The non-integrability of our system, a BEC in an optical
lattice, is indicated by the dynamical instability found in this
system both theoretically~\cite{Wu2000Landau,XuYongP} and
experimentally~\cite{Fallani2004PRL}.

The article is organized as follows. In Section II, we briefly describe our
experimental setup and report that the evolution of our BEC system undergoes
three stages.  In Section III, we give a detailed description and an analysis of the second
stage, where the quantum equilibrium is reached. In Section IV, we describe the
third stage, where the classical thermalization is finally reached. Then we conclude in Section V.


\begin{figure}
\begin{center}
 \includegraphics[width=0.4\textwidth]{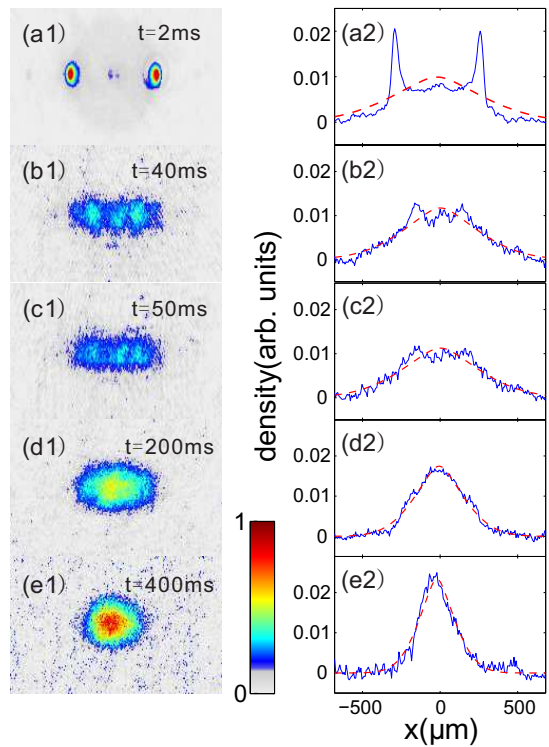}
\end{center}
\caption{(color online) The TOF images (left column) and the
corresponding integrated one dimensional distribution (right column)
at five typical times: $t=2$ ms, $t=40$ ms, $t=50$ ms, $t=200$ ms, and
$t=400$ ms.  The red dashed lines in the right column are the best
thermal distribution fit. $V=8E_\mathrm{r}$.} \label{figure2}
\end{figure}

\section{Experimental setup}
The experimental setup is similar to our previous
work~\cite{Hu2015}. A nearly pure condensate of about $1.5\times
10^5$ $^{87}$Rb atoms is obtained in our hybrid optical-magnetic
trap whose harmonic trapping frequencies are
$(\omega_x,\omega_y,\omega_z)=2\pi\times(28,55,65)$Hz. A one
dimensional optical lattice is formed along the $x$-direction by
retro-reflecting a laser beam with wavelength $\lambda=852$ nm. The
lattice constant is then $a=\lambda/2=426$ nm. The lattice depth is
expressed in units of recoil energy $E_\mathrm{r}=\frac{\hbar^2
k^2}{2m}$ with $k=2\pi/\lambda$. The condensate is quickly loaded
into the p-band (first excited band) of the optical lattice by using
a series of pulsed optical lattices. The pulses are tens of
microseconds wide and consist of two sets whose lattice sites are
shifted in the $\hat{x}$ axis by $a/4$. The details of our method
can be found in Ref.~\cite{Hu2015}. The condensate prepared in such
a way has a narrow width of quasi-momentum around $q=0$. Lots of
interesting physics has been studied both theoretically and
experimentally for a BEC in the
p-band~\cite{Li2011Time,Liu2006Atomic,Kock2016Orbital,M2013Interaction,Wirth2011Evidence}.
In this work we focus on its dynamical relaxation.

We hold the condensate in the p-band for a period of time $t$ up to
several hundreds of milliseconds. Then all the potentials are
switched off and the atom cloud is released. After a 28 ms free
expansion, we take the time-of-flight (TOF) absorption image, which
shows the momentum distribution of the atomic cloud. TOF images at
five typical holding times are shown in Fig. \ref{figure2}.
Initially there are two peaks at $q=\pm\hbar k$, clearly indicates
that the condensate is in the p-band~\cite{Wirth2011Evidence}. As
the evolution goes on, these two peaks begin to disappear and a
central peak with a flat top emerges around tens of milliseconds,
and the distribution stays unchanged for a period of time, an
indication that the quantum equilibrium is reached. At 200 ms, the
central peak is in a familiar Gaussian distribution. At 400 ms, we
not only observe the thermal distribution but also a round cloud
shape that is a signature of classical thermal
equilibrium~\cite{Henn2009}.

We have analyzed these TOF images in detail, which shows that the
whole evolution can be divided into three typical stages. There is
an initial oscillation period roughly before 20 ms and this stage is
characterized by the two prominent Bragg peaks in the images. The
detailed analysis of this stage has been done in our previous
work~\cite{Hu2015}. The second stage follows immediately. In this
stage, the system enters into a stable state, where all the
quantities that we can measure and have measured remain almost
constant. As the system is still in a quantum pure state during this
stage, we call this stage quantum equilibrium stage. After this
stage, oscillations of a different type start the third stage and
they eventually die out. At the third stage, all the TOF images can
be well fit by a Gaussian function. At this final stage the system
is in a mixed state due to inevitable
experimental noise and finally becomes thermalized classically.\\



\section{Quantum equilibrium}
According to von Neumann~\cite{neumann1929,neumann2010} and
others~\cite{reimann2008,han2014}, a non-integrable quantum
many-body system starting from a well-behaved pure state, such as a
Gaussian packet and a Bloch state, will eventually evolve
dynamically into an equilibrium state which looks intuitively rather
random or irregular. As a result, there are two stages of dynamical
evolution. In the first stage, which is usually short and
characterized by a relaxation time, the quantum system undergoes a
certain type of coherent dynamics, which will quickly be destroyed
by dephasing. To see this clearly, let us write the dynamics of a
quantum system in its general form
\begin{equation}
\ket{\psi(t)}=\sum_n c_n e^{-iE_n t/\hbar}\ket{E_n}\,,
\end{equation}
where $\ket{E_n}$ is the system's energy eigenstate with
eigen-energy $E_n$ and coefficient $c_n$ determined by the initial
condition. For a non-integrable quantum system, the structure of its
energy eigenvalues $E_n$ is very similar to the one of a random
matrix~\cite{GutzwillerBook}. As a result, the phases $e^{-iE_n
t/\hbar}$ will quickly be scrambled as $t$ increases. The dephasing
occurs, causing the quantum system to
equilibrate~\cite{neumann1929,neumann2010,reimann2008,han2014}. In
this way, the quantum system enters the second stage, where besides
small fluctuations all the observables become constant.

Note that the above theoretical discussion is for a quantum system ideally isolated from
the environment and the quantum system is still in a pure state even in the second
equilibrium stage. In a real experiment, the quantum system is always coupled to an
environment, which can drive a quantum pure state into a mixed state.
If the coupling is strong, the second quantum equilibrium stage may never happen
as the quantum system can quickly be driven into a mixed state and becomes classical.
When the coupling is weak, the second stage can survive for a period of time before
entering the third stage, where the system evolves into a mixed state
and eventually equilibrate classically.

In our experiment, the coupling to the environment is weak enough
and we have indeed observed all the three stages. We use $t_1$ to
denote the time the system transition from the first to the second
stage and $t_2$ the time the third stage begins. In the first stage
coherent oscillations with decay amplitude are observed along with
other dynamical features. We have analyzed this stage in detail in
Ref.~\cite{Hu2015}. We shall focus on the second and the third
stages.

We attempt to characterize the TOF images quantitatively by
using the aspect ratio of the cloud first. We calculate, $r_\mathrm{x}$ and
$r_\mathrm{y}$, the full widths at half maximum (FWHM) of the
atomic cloud in the TOF images along the $x$ and $y$ axes,
respectively. The FWHM in each direction is obtained by integrating
the two-dimensional atom distribution in the perpendicular direction to get a one-dimensional momentum
distribution and then counting pixels with atom number higher than
half of the maximum value. The aspect ratio is then
$r=r_\mathrm{x}/r_\mathrm{y}$. We have plotted how the aspect ratio
changes with time for the case of $V=8E_\mathrm{r}$ in
Fig.\ref{figure1}(a), where each point is the average over five
experiments with error bars given by the standard deviation. It is
clear that there is a plateau during which the aspect ratio remains largely
constant. Note that the ratio in this plateau is about 3, which is
far away from 1, the aspect ratio of a thermal cloud (see
Fig.\ref{figure2}(e1)). This means that the cloud is still quantum
during the plateau.

\begin{figure}[h]
\begin{center}
 \includegraphics[width=0.4\textwidth]{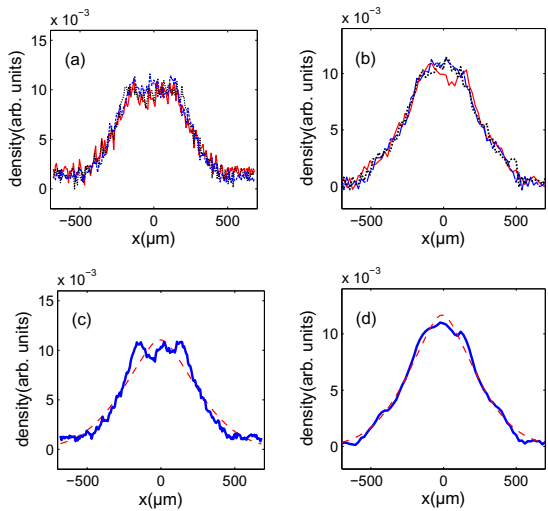}
\end{center}
\caption{(color online) The momentum distribution at the beginning
(red solid line), in the middle (black dotted line), and at the end
(blue dashed line) of the quantum equilibrium plateau for (a)
8$E_\mathrm{r}$ and (b)14$E_\mathrm{r}$. The momentum distribution
averaged over the entire equilibrium plateau is shown in (c) for
8$E_\mathrm{r}$ and (d) for 14$E_\mathrm{r}$ with solid blue lines
while the thermal fitting is shown with red dashed lines.}
\label{figure3}
\end{figure}

In Fig.\ref{figure3} we have plotted the momentum distributions at
different times of this plateau. To demonstrate that the
distribution changes little over the plateau, the momentum
distributions at the beginning, in the middle, and at the end of the
plateau are given in Figs.~\ref{figure3}(a) and (b) for the lattice
depths of $8E_\mathrm{r}$ and $14E_\mathrm{r}$, respectively.
Besides some small fluctuations, the distributions at different
times are clearly the same. To reduce the background noise, each
line is the average over five experiments with the same holding
time. Furthermore, we have plotted in Figs.~\ref{figure3}(c) and (d)
the averaged momentum distribution over the entire plateau for
$V=8E_\mathrm{r}$ and $V=14E_\mathrm{r}$, respectively. Similar to
the distributions at individual times, these two averaged
distributions have a flat top and can not be fit well with a
Gaussian. All these features strongly indicate that the system has
reached a quantum equilibrium, where the distribution has a rather
flat top and can not be fitted with the thermal Gaussian
distribution.

The two transition times $t_1$ and $t_2$ can be extracted. Our
criterion considers both the fluctuation of the aspect ratio $r$ and
the momentum distribution. We first determine $t_1$ and $t_2$ by
requiring the aspect ratio $r$ fluctuates less than $10\%$ of the average value. The
average value is also modified when a new point is included. For
lattices with strength bigger than $14E_\mathrm{r}$, the criterion
is changed to $20\%$ when the time span is longer than 20 ms, as the
fluctuation of experimental result is larger for higher lattice
depth. Then time $t_2$ is adjusted by looking into the momentum
distribution. For example, in the case of $V=8E_\mathrm{r}$, we have
$t_1=36$ ms and $t_2=62$ ms by just considering the fluctuation of
$r$. However, when examining the momentum distribution, we find that
the center of momentum distribution begins to change gradually from
a flat top to a Gaussian one starting at $t=52$ ms. As a result, we set
$t_2=52$ ms, instead of $62$ ms. For other lattice depths, this
phenomenon is also existed, and each $t_2$ is roughly subtracted by
10 ms after considering this effect.

These two times are plotted in Fig.\ref{figure1}(b) as a function of
lattice strength $V$ with $t_1$ as blue filled squares and $t_2$ as
red open squares. The transition time $t_1$ is seen decreased with
the lattice strength, indicating that the stronger lattice renders
the cloud to quantum equilibrium faster. This is quite reasonable:
If we use the dynamical instability to characterize how strong the
chaos of the system is, it is known in literature that a BEC in an
optical lattice is more chaotic for stronger
lattice~\cite{Wu2000Landau}. Usually more chaotic systems have
shorter relaxation times.

The other transition time $t_2$ remains almost constant around 52
ms. This observation is also consistent with our basic
understanding. Our experimental system is weakly coupled to an
environment, which includes thermal atoms~\cite{Griffin},
fluctuations of laser field~\cite{Pichler}, inelastic scattering of
photons~\cite{Zoller,Gerbier}. These noises can eventually destroy
the ``quantumness" of the system and turn it from a pure state to a
mixed state. As this coupling to the environment is insensitive to
the details of the BEC system, one expects that $t_2$ should be
independent of the lattice strength. This is indeed what we have
observed. Our estimation shows that the collision from the thermal
atoms dominates the environment effects and the relaxation time due
to the thermal collision is about 83 ms~\cite{Griffin}, which is
consistent with $t_2$. While the effect of both noise in intensity
of laser field and the inelastic scattering of photons would induce
a decay in a timescale of several seconds.

Another strong evidence that our system is still in a pure quantum
state in the second stage comes directly from our own experiment in
Ref.~\cite{ZhongKai}. The experimental setup is the same.  The only
difference is that the BEC is loaded into the f-band in
Ref.~\cite{ZhongKai}, where quantum coherent oscillations similar to
Bloch oscillations were observed up to 60 ms. Quantum equilibration
was not observed in Ref.~\cite{ZhongKai}. The reason is that the
kinetic energy dominates in the higher bands and the interaction can
be ignored so that the system is integrable.

According to von Neumann~\cite{neumann1929,neumann2010}, the
equilibrium state we observed in the second stage is caused by the
non-integrability of the system. Specifically for our BEC system,
the non-integrability comes from the interaction between atoms. The
collisions between the atoms can deplete the p-band and render the
atoms to the s-band and higher bands or lateral motion. The reversal
process can also occur. At the end, these two processes can balance
out and our BEC system reaches equilibrium.

The details of the quantum dynamical evolution in the second stage
can in principle be described by the many-body Schr\"odinger
equation. However, at present there is no tractable way to solve
this equation for our system, which is initially loaded to the $q=0$
state of the p-band, as this state is not even a local energy
minimum. The mean-field Gross-Pitaevskii equation can only describe
the early moments of the dynamics before the Ehrenfest time (much
shorter than $t_1$)~\cite{Ehrenfest} due to the existence of
dynamical instability~\cite{XuYongP}.


\section{Classical thermalization}

\begin{figure}
\begin{center}
 \includegraphics[width=0.4\textwidth]{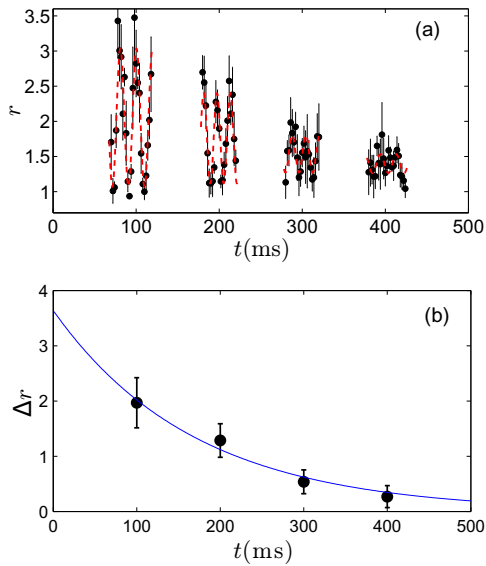}
\end{center}
\caption{(color online)(a) Oscillations of the aspect ratio $r$. The
dashed line is a theoretical fitting. The oscillation frequency is
about twice the trapping frequency along the direction of optical
lattice. (b) Amplitudes of the oscillations. The solid line is an
exponential fit. $V=8E_\mathrm{r}$.} \label{figure4}
\end{figure}

The third stage
of the evolution starts around $60$ ms, and is characterized by its
Gaussian momentum distribution. Interestingly, the aspect ratio of
the cloud starts to oscillate again in this stage but with a
different frequency. These oscillations last for a long time until
the system eventually reaches the classical equilibrium around a
few hundreds of milliseconds. As the cloud with $r_y$ is roughly
constant, the width $r_x$ oscillates in an
identical fashion with $r$. The oscillations of the aspect ratio $r$
for the case of $V=8E_\mathrm{r}$ are shown in
Fig.~\ref{figure4}(a). They can be well fitted by slowly-decaying
sine functions. Through the fitting, we find the oscillation
frequency is $\omega=2\pi\times55.3\pm0.49$Hz. This is
approximately twice of the trapping frequency
$\omega_x=2\pi\times28$Hz. This frequency doubling is independent of
the lattice strength.

This oscillation phenomenon is of classical nature and can be explained as
follows. A gas in a harmonic potential can be regarded as a gas of
harmonic oscillators with the same frequency. As a result, according
to the law of equipartition of energy, its kinetic energy must be
equal to its potential energy at equilibrium. In our experiment,
when the quantum gas loses its coherence and becomes classical at
time $t_2$, it is yet to reach equilibrium with the trapping
harmonic potential. As each atom in the gas oscillates in the
trapping potential, the momentum distribution of this gas will
oscillate accordingly with a doubled frequency. The reason is that
after half of the harmonic oscillation cycle, each atom would change
the direction of its momentum in $\hat{x}$ axis while maintaining
the magnitude. Due to the symmetry of the system, the overall momentum
distribution would have restored the initial state.

As the atomic cloud is not ideally isolated in experiments and the
trap is not perfectly harmonic, it will
eventually equilibrate with the trapping potential. This is
demonstrated by the damping of oscillation amplitude of the aspect
ratio  $\Delta r$ as shown in Fig. \ref{figure4}(b). A numerical
fitting shows that the damping follows an exponential form
$A_0e^{-t/t_d}$ with $t_d=169.8\pm28.7$ ms. Such a long
relaxation time shows that the system is well isolated and is an
indirect indication that the equilibration observed in the plateau
is of quantum nature. For lattice depth of 8$E_\mathrm{r}$ the
aspect ratio of the atomic cloud becomes constant around 400 ms.
For a deeper lattice, the system would reach thermal equilibrium
faster. \\

\section{Conclusion} In sum, we have studied
experimentally the dynamical relaxation of a non-integrable quantum
system by loading a BEC into the second band of the optical lattice.
By following its time evolution, we have observed a quantum
equilibrium state, which is characterized by a constant non-Gaussian
momentum distribution. Our study here has presented a preliminary
experimental test of the two fundamental theorems proved by von
Neumann in his pioneering work~\cite{neumann1929,neumann2010}. Much
more is needed to clarify many aspects of this dynamical relaxation.
For example, what else can we measure to characterize the quantum
equilibrium? And ultimately, can we measure the quantum entropies
for quantum pure states defined by von
Neumann~\cite{neumann1929,neumann2010} or in Ref.~\cite{han2014}.

\acknowledgements
We acknowledge helpful discussion with Hongwei Xiong.
This work is supported by NSFC (Grants N0.61475007, No. 1274024, No.
11334001, and No. 1429402), and the National Basic Research Program of China
(Grants No. 2013CB921903 and No. 2012CB921300).
\bibliography{qequilibration}

\end{document}